\def\thx{\theta_{12}}
\def\sh{s_h}
\mathchardef \bigopen="1314
\mathchardef\bigclose="1315
\mathchardef \paropen="1312
\mathchardef\parclose="1313
\def\uglu{\hskip 0pt plus 1fil minus 1fil}
\def\uglux{\hskip 0pt plus .75fil minus .75fil}
\def\slashed#1{\setbox200=\hbox{$ #1 $}
    \hbox{\box200 \hskip -\wd200 \hbox to \wd200 {\uglu $/$ \uglux}}}

\def\slp{\slashed p}
\def\slq{\slashed q}

\def\slam{\slashed \lambda}
\def\slee{\slashed {\cal E}}
\def\fanA{{\cal A}}
\def\fanB{{\cal B}}

\def\fanD{{\cal D}}
\def\fanE{{\cal E}}
\def\fanF{{\cal F}}

\def\fanL{{\cal L}}
\def\fanM{{\cal M}}
\def\fanN{{\cal N}}

\def\fanR{{\cal R}}
\def\fanS{{\cal S}}

\def\fanU{{\cal U}}

\def\fanZ{{\cal Z}}

\newskip\humongous \humongous=0pt plus 1000pt minus 1000pt

\newif\ifdtup

\documentstyle[12pt]{article}
\begin{document}
\begin{titlepage}
\begin{center}
\begin{flushright}
SLAC-PUB-6425\\
January 1994\\
T
\end{flushright}
\vfill
\large\bf
Influence of Kaonic Resonances on the CP Violation in
$B\to K^*\gamma$
Like Processes*
\normalsize
\rm
\bigskip\bigskip

D. Atwood$^{\rm a}$ and A. Soni$^{\rm b}$ \\
\bigskip
\end{center}

\begin{flushleft}

a) Department of Physics, SLAC, Stanford University, Stanford, CA\ \ 94309, USA
b) Department of Physics, Brookhaven National Laboratory, Upton, NY\ \ 11973,
US

\end{flushleft}
\bigskip\bigskip

\bigskip

\vfill

\begin{center}
Submitted to {\it Zeitschrift f\"ur Physik C}
\end{center}

\vfill

\hrule
\vspace{5 pt}
\noindent
* Work supported by the Department of Energy Contracts
DE-AC03-76SF00515 (SLAC) and DE-AC02-76CH0016 (BNL) and an
SSC Fellowship (for David Atwood).
\end{titlepage}

\small
\begin{center}
{\bf Abstract}
\end{center}

We consider CP violating effects
in
decays of the type
$B\rightarrow k_i\gamma \rightarrow K\pi\gamma$,
$K^*\pi\gamma$
and $K\rho\gamma$,
where $k_i$ represents a strange meson resonance.
We include in our calculations five of the low-lying
resonances with quantum numbers ($J^P$) $1^-$, $1^+$ and
$2^+$.
At the quark level these decays are driven by the penguin graph as
well as tree graphs. CP violation arises in the Standard Model
due to the difference in the CKM
phase between these graphs.
We model
the final state interaction of the hadronic system
using the low lying $k_i$ resonances.
Bound state effects are incorporated by using ideas based on the
model of Grinstein et al
and also in another bound state model
(somewhat similar to the model of Wirbel et al)
which we construct
designed to take into account relativistic effects better.
In these models we find that radiative decays of B mesons
give rise to four of the five kaonic resonances at about
$1$ to $7 \%$ of the inclusive $b \rightarrow s \gamma$ rate.
Furthermore, in both bound state models we find
that the probability of formation of the other three
higher resonances is roughly the same as that of $K^*(892)$,
which was recently seen at CLEO.
In addition to the
partial rate asymmetry which arises due to interference
between
resonances of the same quantum numbers,
we show how
interference between resonances
of the same parity, and also between resonances of
the opposite parity, result in two different types
of energy asymmetries.
CP differential
asymmetries at the level of a few percent seem possible.
We thus obtain CP violating
distributions which
may be
observed in a sample of
about $10^9$\relax $B^\pm$
mesons.
For concreteness in this paper we deal with only charged B's, neutral
B's will be dealt with separately.
\normalsize

\eject

\section{Introduction}

The recent observation \cite{ref_cleo}   of the long
awaited process $B\to K^* \gamma$ gave another reminder of the possible
richness
of the physics which could be observed at $B$ factories.

In this paper we will consider the generalizations of  the above decay,
{\it i.e.}\ we consider the  decays
$B\to k_i\gamma$ where the $k_i$ denote  excited $K$ meson states
or resonances.
In particular, we wish to investigate if widths of resonances can be
used to enhance CP violation effects in B-decays;
their importance in the context of the top quark
has been emphasized in the past few years\cite{ehs,ehs2}.
The key difference is that in the case of B-decays
the resonances are
strongly interacting so that
width to mass ratio is much larger than
was the case for the top decays wherein
the resonances are electroweak in character.

In the case at hand, namely $B \rightarrow k_i \gamma$,
$k_i$ must have $J\ge 1$; so,
in particular we will
focus on the five lowest lying such states.  We will denote these states
$k_0$,   $k_1$,   $k_2$,   $k_3$ and  $k_4$.   The first such state  which
we write $k_0$ is  $K^*(892)$; it has quantum numbers $1^-$.  In a
constituent quark model it is a $\bar u s$ (or $\bar d s$) quarkonium
state $^3S_1$.
The next two states that
we are interested in
we will denote as $k_1$ and $k_2$. They both have quantum numbers
$1^+$ and
in the notation of
\cite{part_data} are
written as $K_1(1270)$ and $K_1(1400)$.
In the quark model they should correspond to
mixtures of
the states
$^1P_1$ and
$^3P_1$.
For these pure quark model states we will use the notation
$\hat k_1 = {}^1P_1$
and
$\hat k_2 = {}^3P_1$ and so, following \cite{kspec}
the physical states are
related to these by a
mixing angle $\thx$:
\begin{eqnarray}
k_1&=& \cos\thx\ \hat k_1-\sin\thx\ \hat k_2\nonumber\\
k_2&=& \sin\thx\ \hat k_1+\cos\thx\ \hat k_2.
\end{eqnarray}
Though strictly speaking the states $\hat k_1$ and $\hat k_2$ are
not eigenstates of charge conjugation, in the literature they are
sometimes referred to as $1^{+-}$ and $1^{++}$ states since they are
related by $SU(3)$ to the $b_1(1235)$
and $a_1(1260)$ mesons with those quantum
numbers.
In fact the mixing here has been observed to be close to maximal; in
\cite{kspec} $\thx$ is experimentally determined
to be $56\pm 3^\circ$.

The  $2^3S_1$ state
$K^*(1410)$
will be denoted as
$k_3$.
In principle $k_3$,
which is a
radially excited $2^3S_1$ state,
could also mix
with $k_0$ however since the masses are so far apart it is unlikely
that this mixing is large.
We will therefore ignore this mixing in our calculation.
We also consider the
$K_2(1430)$ state with quantum numbers $2^+$ which
will be designated
$k_4$.
$D$ wave states have also been observed around 1700 MeV but
we will not  include these in our analysis.
We have summarized some of the
known properties \cite{part_data} of these states in Table 1.

Consider the two decays
$B\to k_i\gamma ;\  k_j \gamma$
followed by decays of
$k_i$ and $k_j$
decay to a common hadronic final state $XY$.
If the two
channels
have
different CP phases then CP violation could manifest
in the momentum distribution of the  mesons making up $XY$. Thus, for
example, the energy distribution of one of the particles in $XY$ may be
different for $B$ decay
compared to that in
$\bar B$ decay.  If the quantum
numbers of the two states are the same, as is the case for $k_1$ versus
$k_2$ and $k_0$ versus $k_3$ then there also exists the possibility of a
partial rate asymmetry (PRA) between $B$ decay
and $\bar B$ decay. Of course
in the case of neutral  $B$ decay one must also consider these phenomena
in the context of  $B-\bar B$ oscillations
\cite{atw_soni_prog}.

We find that the largest effect occurs in the case of the final
state $K^*\pi$. In this case we estimate that a difference between the
distribution of the $K^*$ in the decays of $B^-$
versus $B^+$ may well be
observable with about $5\times 10^8$ $B^\pm$ decays.
The
final state $K\pi$
seems to
require about $5\times10^9$ $B^\pm$
decays.
In passing, we mention that, whereas we are focussing on influence of
resonances to CP violation in radiative decays of B {\it mesons}
to {\it exclusive} channels, CP violation in {\it inclusive},
radiative decays of the
b {\it quark} has been examined by Soares \cite{soares}.
The two approaches are therefore somewhat complementary.

The rest of the paper will proceed as follows: In section 2 we will
explain how it could happen that the different $k_i$ decays could
acquire different CP phases, in particular in the
Standard Model (SM);
we will also estimate the magnitude these phases could have.
In making these estimates, for incorporating bound states effects,
we will use ideas based on references \cite{isgur,wsb_ref}.
In section 3 we will explain how these CP
phases will
give rise to
CP violating kinematic distributions
as well as partial rate asymmetries.
In that section
we also estimate the magnitude of various
asymmetries to be expected.

\eject

\section{Basic Processes}

Any CP violating phase which enters into the process
$B\to k_i\gamma$ must have its origin  either in the electroweak
physics which drives the process or in physics beyond the standard
model.  In the standard model, three classes of quark graphs may
contribute as
shown in Figure 1. Figure 1a shows the penguin graph for the  quark
level process $b\to s\gamma$ which has been studied  extensively
\cite{penguina,penguinb}.
Figure 1(b) shows an annihilation
process which is operative  only in the case of $B^\pm$. This
process will  give a  $\gamma k_i$  state if the two quarks coalesce
into the appropriate $k_i$
state.
Figure 1(c) shows a spectator process which
could give rise to  $\gamma k_i$  if the four quarks shown should
coalesce into
two mesonic states and thence to
a  $k_i$ state.

Of course to have observable CP violating effects one
need not only have a CP violating phase but there must be the interference
of processes with different CP phases. In the standard model such a phase
is given by the CKM matrix ($V$).
Introducing the standard Wolfenstein \cite{wolf}
parameterization of the CKM
matrix
\cite{paschos}:
\begin{equation}
V =  \left(
\begin{array}{ccc}
1-{\lambda^2\over 2}  & \lambda             &
A\lambda^3\sigma e^{-i\delta}\\
-\lambda              & 1-{\lambda^2\over 2}&
A\lambda^2                  \\
A\lambda^3(1-\sigma e^{i\delta})& -A\lambda^2 &
1                       \\
\end{array} \right)
\label{vkmdef}
\end{equation}
where $\lambda = sin\theta_{cabibbo} = 0.22$.
The amplitude for the penguin graphs \cite{penguina,penguinb}
will be proportional to the
quantity:
\begin{equation}
\sum_{i=u,c,t} V^*_{is} V_{ib} F(m_i)
\end{equation}
where the form of $F$ for the graph in Figure 1a is given in
\cite{penguina} and the QCD corrected form is given in \cite{penguinb}.
Note that the above sum will be dominated by the $i=c$
and $i=t$ terms, hence from
(2) we see that
in this parameterization there is little phase in the penguin
amplitude.
Both the spectator and the annihilation graphs will,
on the other hand, be proportional to the product

\begin{equation}
V^*_{us} V_{ub}
=A\lambda^4\sigma e^{-i\delta}
\end{equation}

\noindent which has a phase of precisely $-\delta$ in this notation.
Thus interference between the penguin graph and either the spectator or
the annihilation graphs potentially can produce observable CP violating
effects.

To see how this comes about consider, for instance, the two decay channels
$B\to k_2\gamma \to \pi K^* \gamma$ and $B\to k_4\gamma \to \pi K^*
\gamma$.  The final states are the same, hence there could be
interference effects  between them. If there is a difference in CP phase
as well as a difference in  the strong interaction phase
({\it i.e.}\ CP
conserving phase) , there could be a difference in the energy
or other
distributions between $B$ and $\bar B$ decay which would clearly
violate
CP\null.  We will assume that near these resonances the strong
interaction phase is dominated by  Breit-Wigner forms for the
$k_i$
propagators.

In order to calculate the CP phase for the decay
$B \rightarrow k_i \gamma$
we will thus need
to estimate the
relative contribution of each of the three classes of graphs to the five
$k_i$ resonances that
we are considering.

In the case of the annihilation and spectator graphs it will be useful
to compare these rates with
$\Gamma_{bus}^0$,
the tree level inclusive process
$b\rightarrow u\bar u s$ given by:
\begin{equation}
\Gamma_{bus}^0=
3{G_F^2 m_b^5\over 192\pi^3}
|V_{ub}V_{us}^*|^2
=
{3 f^{-1}({m_c^2\over m_b^2})}
\left |
{V_{ub}V_{us}^*\over V_{cb}}\right |^2
\Gamma(b\rightarrow e \nu c)
\label{gam0def}
\end{equation}
where
\begin{equation}
f(x)=1-8x+8x^3-x^4-12x^2\log x
\end{equation}
is the phase space factor defined in
\cite{gin_gla_wise}; in this case
$f(m_c^2/m_b^2)=0.46$, where we are using
$m_c=1.5GeV$ and $m_b=4.6GeV$.

For a given B decay, via the spectator graph,
to the final state $X$ we define
\begin{equation}
r(X)={\Gamma(X)\over \Gamma_{bus}^0}.
\label{r_def}
\end{equation}

\subsection{Penguin Graph}

The dominant graph as we shall see is likely to be the penguin process
depicted in Figure 1(a). In this case the $k_i$ system is formed by the
merging of the spectator and the $s$ quark.
In order to calculate the probability of the formation of each of the
resonances, one needs a model for the bound state effects involved.
For this purpose we will consider potential models
\cite{isgur,wsb_ref,desh_lo_tram,ber_hsi_soni}.

In general such a
model will relate meson level amplitude
$\fanM_m$ to the quark level amplitude $\fanM_q$ according to the formula:
\begin{equation}
\fanM_m=\int \fanM_q \Phi^B(P) \Phi^k_i(\hat P) dP
\label{eq_master}
\end{equation}
where the meson wave functions $\Phi^B$ and $\Phi^k_i$ are
functions of quark momenta and spins with the correct quantum numbers
to form the indicated mesons.
Here $P$ represents the momentum of the $b$ quark in
the  rest frame of the $B$ meson
and
$\hat P$ represents the momentum of the $s$ quark
in the frame of the
$k_i$.
The exact details of how the integral is constructed
will depend on the specific model.
We will construct two such models which we will designate A and B
in order to get a feel for the accuracy of the predictions.

Model A is similar to the one used in  \cite{desh_lo_tram} and is
based on the quark model of Grinstein {\it et al}. \cite{isgur} which
has been quite successful in semileptonic charm and bottom decays.
Model B is
constructed to take relativistic effects better and is
based roughly on the ideas of Wirbel {\it et al}.
\cite{wsb_ref}. For model A, non-relativistic kinematics is used for
the quarks and  $P$ is taken to be the 3 momentum of the $b$-quark in
the $B$-meson rest frame and  $\hat P$ is given by

\begin{equation}
\hat P=\vec P-x_u \vec K
\end{equation}
\noindent where $x_u=m_u/(m_u+m_b)$ and $\vec K$ is the momentum of
the $k_i$ meson
in the B rest frame.

The wave functions for the various meson states are approximated by
harmonic oscillator functions. Thus the
momentum dependent part of the wave functions are:
\begin{eqnarray}
\Phi_B&=&
\pi^{-{3\over 4}}\beta_B^{-{3\over 2}}
e^{-{P^2\over 2\beta_B^2}} \nonumber \\
\Phi_{1S}&=&
\pi^{-{3\over 4}}\beta_S^{-{3\over 2}}
e^{-{P^2\over 2\beta_S^2}} \nonumber \\
\Phi_{1P(0)}&=&
\pi^{-{3\over 4}}\beta_P^{-{3\over 2}}
{\sqrt{2}P_z\over \beta_P}
e^{-{P^2\over 2\beta_P^2}} \nonumber \\
\Phi_{1P(\pm 1)}&=&
\pi^{-{3\over 4}}\beta_P^{-{3\over 2}}
{(P_x\mp i P_y)\over \beta_P}
e^{-{P^2\over 2\beta_P^2}} \nonumber \\
\Phi_{2S}&=&\sqrt{3\over 2}
\pi^{-{3\over 4}}\beta_{S}^{-{3\over 2}}
({2P^2\over 3\beta_{S}^2}-1)
e^{-{P^2\over 2\beta_{S}^2}}
\end{eqnarray}
Here $\Phi_{nS}$ represents the $n$th excited $S$-wave state,
$\Phi_{nP(m)}$ represents the  $n\ $th excited $P$-wave state with
angular momentum projection $m$ and $\phi_B$ is the wave function for
the B meson.
The constants $\beta_i$ are determined from the
potential model using a
variational method.
The values
obtained in references
\cite{desh_lo_tram,isgur}
are

\begin{equation}
\beta_B=0.41\mbox{ GeV}\ \ \ \
\beta_S=0.34\mbox{ GeV}\ \ \ \
\beta_P=0.30\mbox{ GeV}
\end{equation}

\noindent where they use the constituent masses

\begin{equation}
m_u=m_d=0.33\mbox{ GeV}\ \ \ \
m_s=0.55\mbox{ GeV}\ \ \ \
m_b=5.12\mbox{ GeV} .
\end{equation}

With the use of the above momentum distributions above we can specify the
spin-dependent part of the wave
function for
the B-meson and for the $k_i$ with $J_z=-1$:

\begin{eqnarray}
|B>&=&{1\over\sqrt{2}}
\Phi_{B}
\left(b(\uparrow)\bar u(\downarrow)+
b(\downarrow)\bar u(\uparrow)\right) \nonumber\\
|k_0>&=&\Phi_{1S}\  s(\downarrow)\bar u(\downarrow) \nonumber\\
|\hat k_1>&=&{1\over\sqrt{2}}
\Phi_{1P(-1)}\
\left(s(\uparrow)\bar u(\downarrow)+
s(\downarrow)\bar u(\uparrow)\right) \nonumber\\
|\hat k_2>&=&
{1\over\sqrt{2}}\Phi_{1P(0)}\ s(\downarrow)\bar u(\downarrow)
-{1\over 2}\Phi_{1P(-1)}\ (s(\downarrow)\bar u(\uparrow)
-s(\uparrow)\bar u(\downarrow)) \nonumber\\
|k_3>&=&\Phi_{2S}\  s(\downarrow)\bar u(\downarrow) \nonumber\\
|k_4>&=&
{1\over\sqrt{2}}\Phi_{1P(0)}\ s(\downarrow)\bar u(\downarrow)
+{1\over 2}\Phi_{1P(-1)}\ (s(\downarrow)\bar u(\uparrow)
-s(\uparrow)\bar u(\downarrow))
\label{wavedef}
\end{eqnarray}
Note that for the $1^+$ states we use the quark model basis
$\{\hat k_1,\hat k_2\}$.

At the quark level the $bs\gamma$  coupling is

\begin{equation}
{\rm a}\ \bar s \  \sigma^{\mu\nu}q_\mu (P_R+{m_s\over m_b}P_L)\  b
\end{equation}

\noindent where $P_R={1\over 2}(1+\gamma_5)$ and $P_L={1\over
2}(1-\gamma_5)$ and $a$ is given in \cite{penguinb}. We will concentrate on the
proportional to $P_R$ which is dominant.
Note that this gives rise to {\it left} handed photons
in the final state.

Let us take the $z$-axis in the direction of the hadronic momentum so
the 4-momentum of the photon is

\begin{equation}
p_\gamma= {m_B^2-m_i^2\over 2} \left[
\begin{array}{c} +1\cr
0\cr
0\cr
-1\cr \end{array}
\right ]
\end{equation}

\noindent and we take the 3-momentum of the $b$-quark to be $\vec P$.
Expanding the amplitude for small $P_x, P_y$ we obtain:

\begin{eqnarray}
\fanM_{++}&=&-{\rm a}\sqrt{8}(m_b-m_s)P_-           \nonumber\\
\fanM_{+-}&=&0                                        \nonumber\\
\fanM_{-+}&=&-{\rm a}\sqrt{8}(m_b^2-m_s^2)            \nonumber\\
\fanM_{--}&=&+{\rm a}\sqrt{2}m_b^{-1}(m_b^2-m_s^2)P_-
\label{eq_amplitude}
\end{eqnarray}

\noindent where $\fanM_{ij}$ is the amplitude for $b$ quark
with spin
projection $S_z=j/2$  going to $s$-quark
with spin projection $S_z=i/2$. The quantity
$P_\pm=P_x\pm iP_y$.

Using the above expansion we can obtain analytic expressions
for the  meson amplitudes:

\begin{eqnarray}
\fanM(k_0)&=&-2{\rm a}
\left( \hat\beta^2\over\beta_B\beta_S   \right )^{3\over 2}
(m_b^2-m_s^2) e^{-\Delta_S} \nonumber\\
\fanM(\hat k_1)&=& -\sqrt{1\over 2}{\rm a}
{\hat\beta^2\over m_b\beta_P}
\left( \hat\beta^2\over\beta_B\beta_P \right )^{3\over 2}
(m_b-m_s)^2 e^{-\Delta_P} \nonumber\\
\fanM(\hat k_2)&=& -{\rm a}\bigopen \left( \hat\beta^2\over\beta_B\beta_P
\right )^{5\over 2} \beta_B^{-1}x_u
\left(m_B^2-m_i^2\over 2 m_B\right ) (m_b^2-m_s^2) \nonumber\\
&& + {1\over 2} \left( \hat\beta^2\over\beta_B\beta_P
\right )^{3\over 2} {\hat \beta^2\over \beta_P m_b}
(m_b-m_s)(3m_b+m_s) \bigclose e^{-\Delta_P} \nonumber\\
\fanM(k_3)&=&-{\rm a}\left( {2\over 3}\right)^{1\over 2}
\left( \hat\beta^2\over\beta_B\beta_S   \right )^{3\over 2}
{3(\beta_B^4-\beta_S^4)+2\beta_S^2(x_uK)^2\over
(\beta_B^2+\beta_S^2)^2} (m_b^2-m_s^2) e^{-\Delta_S} \nonumber\\
\fanM(k_4)&=& -{\rm a}\bigopen \left(  \hat\beta^2\over\beta_B\beta_P
\right )^{5\over 2} \beta_B^{-1}x_u \left(m_B^2-m_i^2\over 2 m_B\right )
(m_b^2-m_s^2) \nonumber\\
&& - {1\over 2} \left(  \hat\beta^2\over\beta_B\beta_P
\right )^{3\over 2} {\hat \beta^2\over \beta_P m_b}
(m_b-m_s)(3m_b+m_s) \bigclose e^{-\Delta_P}
\label{penampli}
\end{eqnarray}
where

\begin{equation}
\Delta={x_u^2(m_B^2-m_i^2)^2\over 2 m_B^2(\beta_B^2+\beta_i^2)}
\ \ \ \
\hat\beta_i=
\sqrt{  2\beta_B^2\beta_i^2\over\beta_B^2+\beta_i^2}
\end{equation}

We can now calculate the ratio
\begin{equation}
R_i=  { \Gamma(B\to k_i\gamma) \over \Gamma(b\to s\gamma) }.
\label{R_def}
\end{equation}
The results are shown in Table 2.
Note
that our results are slightly different than those obtained in
\cite{desh_lo_tram}  because here we expanded the matrix element
only
to first order in $\vec P$.

One thing which is worrisome about the model constructed in this way
is that the transformation from $P$ to $\hat P$ is not relativistic.
The velocity of the mesons however is,  since
$v\over c$ ranges from .85 in the case of $k_2$ to
.95 in the case of $K^*$. This motivates us to consider a modification
which respects relativity.

In model B,
we consider wave functions
$\Psi_i$ which are functions
of 4-momentum and are related to $\Phi_i$ by

\begin{equation}
\Psi_i(P)= N_i \Phi_i(\vec P)  \sqrt{E(m_i-E)\over m_i}
e^{-{(E-E_0)^2\over 2\beta_i^2}}.
\end{equation}

\noindent In this equation
$P\equiv (E,\vec P)$ is the 4-momentum of the quark in the meson $i$ and

\begin{equation}
E_0={m_i^2+m_q^2-m_{\bar q}^2\over 2 m_i}
\end{equation}

\noindent $m_q$ and $m_{\bar q}$ being the constituent masses.
We define the wave function to be 0 if $E$ is outside of the physical range
$0\leq E\leq m_i$
and the normalization constant $N_i$ is determined by the condition

\begin{equation}
\int_{0\leq E\leq m_i} ||\Psi(P)||^2 d^4P\ =\ 1.
\end{equation}

\noindent For the form of the $E$ dependent part of the wave function
we are motivated by the similar form in \cite{wsb_ref}.

Thus, in the reaction $B\to k_i\gamma$  the relation between  $\hat P$
and $P$ becomes, $\hat P = L(P-P_\gamma)$ where

\begin{equation} L=
\left[ \begin{array}{cccc} \gamma   &0&0&-\beta\gamma\\ 0&1&0&0\\
0&0&1&0\\ -\beta\gamma &0&0&\gamma
\end{array} \right ]
\end{equation}

\noindent is the Lorentz boost into the rest
frame of $k_i$. We still use
the quark level amplitudes expanded in $P_x$ and $P_y$ in equation
(\ref{eq_amplitude}) and substitute them into equation
(\ref{eq_master}). Thereby we obtain the results in Table 2  labeled
model B.  Note that these results are somewhat smaller,
compared to those from model A,
particularly
in the case of $k_2$, $k_3$ and $k_4$.

\subsection{ Four Quark Hamiltonian for  B Decays}

The four quark couplings involved in
$\Delta S = 1$
charmless $B^\pm$
decay is
given at tree level by $W^\pm$ exchange.
There are however
potentially large QCD corrections which have been calculated using the
renormalization group approach \cite{rgpapers,misiak}. Following these papers,
let us establish a basis of
$\Delta S=1$
operators that may mix together:
\begin{eqnarray}
O_1^{ij}&=&
(\bar   s_\alpha \gamma_\mu P_L b_\alpha)
(\bar q^j_\beta  \gamma_\mu P_L q^i_\beta)
\nonumber\\
O_2^{ij}&=&
(\bar   s_\alpha \gamma_\mu P_L b_\beta)
(\bar q^j_\beta  \gamma_\mu P_L q^i_\alpha)
\nonumber\\
O_3&=&
(\bar   s_\alpha \gamma_\mu P_L b_\alpha)
\sum_k
(\bar q^k_\beta  \gamma_\mu P_L q^k_\beta)
\nonumber\\
O_4&=&
(\bar   s_\alpha \gamma_\mu P_L b_\beta)
\sum_k
(\bar q^k_\beta  \gamma_\mu P_L q^k_\alpha)
\nonumber\\
O_5&=&
(\bar   s_\alpha \gamma_\mu P_L b_\alpha)
\sum_k
(\bar q^k_\beta  \gamma_\mu P_R q^k_\beta)
\nonumber\\
O_6&=&
(\bar   s_\alpha \gamma_\mu P_L b_\beta)
\sum_k
(\bar q^k_\beta  \gamma_\mu P_R q^k_\alpha)
\label{opdef}
\end{eqnarray}
where $i,j\in\{ u,c \}$ and $k\in\{ u,d,c,s,b\}$. $\alpha$ and $\beta$
are color indices.

The effective Hamiltonian may be expanded
in terms of these
operators in
the following way:
\begin{eqnarray}
H_{eff}&=& 2^{3\over 2} G_F
\bigopen
V_{cs}^*V_{cb}(\sum_{i=1,2} c_i^c(\mu) O_i^{cc}
+ \sum_{i=3,\dots,6} c_i^c(\mu) O_i)                     \nonumber\\
&&+
V_{us}^*V_{ub}(\sum_{i=1,2} c_i^u(\mu) O_i^{uu}
+ \sum_{i=3,\dots,6} c_i^u(\mu) O_i)                     \nonumber\\
&&+
V_{us}^*V_{cb}(\sum_{i=1,2} c_i^{uc}(\mu) O_i^{uc})
+
V_{cs}^*V_{ub}(\sum_{i=1,2} c_i^{cu}(\mu) O_i^{cu})
\bigclose
\end{eqnarray}

Following the treatment in \cite{misiak}
to next to leading order in QCD the operators satisfy the following
evolution equations:
\begin{eqnarray}
\mu {d\over d\mu} c_i^q(\mu) &=&
{\alpha_s(\mu)\over 2\pi} \sum_{j=1,\dots,6} c_j^q A_{j,i}
\nonumber\\
\mu {d\over d\mu} c_i^r(\mu) &=&
{\alpha_s(\mu)\over 2\pi} \sum_{j=1,2} c_j^r B_{j,i}
\label{evoleqn}
\end{eqnarray}
where $q\in\{u,c\}$ and $r\in\{uc,cu\}$
and the one loop $\alpha_s$ is given by
\begin{equation}
\alpha_s(\mu)={12\pi\over (33-3f)\log{\mu\over\Lambda_5}}
\end{equation}
where $\Lambda_5$ is the QCD scale for 5 flavors.
The matrix $A$, to the lowest non-trivial
loop order is  \cite{misiak}:
\begin{equation}
A=\left(
\begin{array}{cccccc}
-{3\over N}    &     3      &    0    &    0    &     0    &  0  \\
    3     & -{3\over N}&-{1\over 3N}&{1\over 3}&-{1\over 3N}&{1\over3}\\
0&0&-{11\over 3N}&{11\over 3}&-{2\over 3N}&{2\over 3}\\
0&0& 3-{f\over 3 N} & {f\over 3}-{3\over N}& -{f\over 3 N} & f\over 3\\
0&0&0&0&{3\over N}&-3\\
0&0&-{f\over 3N}&{f\over 3}&-{f\over 3N}&-6c_F+{f\over 3}
\end{array}
\right)
\end{equation}
and the matrix $B$ is:
\begin{equation}
A=\left(
\begin{array}{cc}
-{3\over N}    &     3      \\
    3     & -{3\over N}
\end{array}
\right).
\end{equation}
Here the QCD color factors $N=3$ and $c_F={4\over 3}$.
The number of flavors $f=5$ since we are interested in the evolution
above $m_b$.

Let us define the coefficients of these operators at $\mu=m_W$ to be the tree
level values, thus
\begin{equation}
c_2^u(m_W)=
c_2^c(m_W)=
c_2^{uc}(m_W)=
c_2^{cu}(m_W)=1
\end{equation}
and all the other coeficients are 0.
This implies that
$c^u_i(\mu)=
c^c_i(\mu)=
c^{uc}_i(\mu)=
c^{cu}_i(\mu)=c_i(\mu)$. In particular $c_1$ and $c_2$ may be solved in
a  simple form.
If we write
\begin{equation}
c_1={1\over 2}(c_+-c_-)\ \ \ \
c_2={1\over 2}(c_-+c_+)
\end{equation}
then the solutions
for $c_+$ and $c_-$ are:
\begin{equation}
c_+(\mu)=\left[ {\alpha_s(m_W^2)\over\alpha_s(\mu^2) }
\right]^{6\over 23}
\ \ \ \
c_-(\mu)=\left[ {\alpha_s(m_W^2)\over\alpha_s(\mu^2)}
\right]^{-{12\over 23}}
\end{equation}

The evolution equation (\ref{evoleqn}) is readily integrated
numerically. If we take $\Lambda_5=0.2GeV$ and $m_b=4.7GeV$ then
\begin{equation}
\begin{tabular}{ll}
$c_+=.846$       &
$c_-=1.397$      \\
$c_4+c_3=-0.016$ &
$c_4-c_3=-0.041$ \\
$c_6+c_5=-0.027$ &
$c_6-c_5=-0.043$ \\
\end{tabular}
\label{cresults}
\end{equation}

\subsection{Annihilation Diagram}

In the case of $B^\pm$ decay it is possible to produce a
$\gamma k_i$ state through the annihilation graphs such as the one
depicted in Figure 1(b).

In general such annihilation graphs can be calculated by relating them to
the weak decay constant $f_B$ which
we take to be 180  MeV \cite{ber_lab_soni}, defined so that
$f_\pi = 130MeV$.
Following the calculation in
\cite{sbs_ref} we assume that the
annihilation takes place at $0$ relative momentum. Thus if $\fanU$ is the
amplitude for $b\bar u$ annihilation at 0 relative momentum where we take

\begin{equation}
p_b=x_b P_B\ \ \ \ \  p_u=x_u P_B
\end{equation}

\noindent then the meson level amplitude  can be written as

\begin{equation}
\fanM={1\over4}f_BTr((\slashed{P}_B+m_B)\gamma_5 \fanU)
\ \
{1\over 3} \delta_{ab}
\label{fbdef}
\end{equation}
where $a$ and $b$ are color indices.

Since our goal is to find contributions which interfere with the
penguin graph we must take the photon from the annihilation
graph also to be left handed.
This photon
polarization leads to
the graph where the
photon is radiated from the $b$-quark vanishing.
In addition graphs where
the photon is radiated from the final state vanish in
the limit that the light quark mass goes to zero.
In contrast the graph where the photon is radiated from the initial
$\bar u$  quark is proportional to
$x_u^{-1}f_B$ and so dominates. In fact in the case of $0$ relative
momentum this graph by itself is gauge invariant so it makes sense
to consider it alone.

The four quark operator which we need to extract from the effective
Hamiltonian is
$(\bar   u_\alpha \gamma_\mu P_L b_\alpha)
(\bar s_\beta  \gamma_\mu P_L u_\beta) $.
Note that the color structure is fixed by the constraint that the
initial $\bar u b$ system is a color singlet.
Furthermore
the CP phase is only present in the CKM product
$V_{ub}V_{us}^*$.
Since we obtain CP violating observables by interference of this
process with the penguin graph, henceforth we will extract only the
portion of the amplitude proportional to  $V_{ub}V_{us}^*$.
The total coeficient
for the above operator
after suitable Fierz
transformation is
$\fanD=2^{3\over 2}G_F D V_{ub}V_{us}^*$
where
\begin{equation}
D=-({1\over 3}c_1+c_2+{1\over 3}c_3+c_4)
\label{Ddef}
\end{equation}
Using the numerical results in (\ref{cresults})
we calculate $D=-1.029$.
Operators with the current structure of $O_5$ and $O_6$ do
not contribute
to the photon emission from the initial $\bar u$ quark
as may readily be verified by substitution into
(\ref{fbdef}).
Emission from the final legs is also
suppressed as it does not go like $1/m_u$
or $1/m_s$.

Using equation (\ref{fbdef}) it is straightforward to calculate
the
amplitude to specific spin states which we will denote
$\fanN_{{\cal S},m}$ where
$|{{\cal S},m}>$ is the angular momentum of the $\bar u s $ system
quantized in the $z$-direction:
\begin{eqnarray}
\fanN_{1\ +1}&=&
\fanZ P_-^2
\nonumber\\
\fanN_{1\  0}&=&
-{\fanZ\over \sqrt{2}} (\sqrt{s}+m_s+m_u+2P_z)P_-
\nonumber\\
\fanN_{1\ -1}&=&
+\fanZ (E_u+m_u+P_z)(E_s+m_s+P_z)
\nonumber\\
\fanN_{0\ 0}&=&
-{\fanZ} {(m_s-m_u)(m_s+m_u+\sqrt{s})\over\sqrt{2s}}P_-
\end{eqnarray}
where $s=(p_u+p_s)^2$; in the $\bar u s$ rest frame the 4-momentum of
the s quark is
\begin{equation}
p_s=\left(
\begin{array}{c}
E_s\\
P_x\\
P_y\\
P_z
\end{array}
\right)
\end{equation}
The energy of the $\bar u$ is $E_u=\sqrt{s}-E_s$ and
$P_\pm=P_x\pm i P_y$.
The factor $\fanZ$ is:
\begin{equation}
\fanZ
=
{G_F m_B f_B e_u e V_{ub} V_{us}^*
\over
m_u \sqrt{(E_u+m_u)(E_s+m_s)}}
D
\end{equation}
where $e_u e$ is the charge of the u quark and $D$ is the
coeficient defined in equation \ref{Ddef}.
We now consider two different methods for estimating
the amplitude
for $B\rightarrow \gamma k_i$ from this annihilation process.
First of all we use the nonrelativistic quark model A above and then we
will try to estimate the amplitude in a model independent way.

In terms of a nonrelativistic wave function, the meson level amplitude
${\bf N}_i$ is given by:
\begin{equation}
{\bf N}_i={1\over 2}\pi^{-{3\over 2}} m_i^{-{1\over 2}}
\int \fanN  |k_i> d^3\vec P
\end{equation}
where the wave functions $|k_i>$ are those given in
equation (\ref{wavedef}) and $\vec P$
is the 3-momentum of the s quark in the $k_i$ frame.

We may evaluate this integral analytically
if we use the non-relativistic
approximation
$E_u\approx m_u$ and
$E_s\approx m_s$.

Performing this integral the meson amplitudes thus obtained are:
\begin{eqnarray}
{\bf N}_0&=&
+\fanZ\sqrt{2}\pi^{-{3\over 4}} m_i^{-{1\over 2}} \beta_S^{3\over 2}
(4 m_u m_s + \beta_s^2)
\nonumber\\
\hat{\bf N}_1&=&
-4\fanZ\pi^{-{3\over 4}} m_i^{-{1\over 2}} \beta_P^{5\over 2}
(m_s-m_u)
\nonumber\\
\hat{\bf N}_2&=&
-2\sqrt{2}\fanZ\pi^{-{3\over 4}} m_i^{-{1\over 2}} \beta_P^{5\over 2}
(m_s+m_u)
\nonumber\\
{\bf N}_3&=&
+\fanZ\sqrt{3}\pi^{-{3\over 4}} m_i^{-{1\over 2}} \beta_S^{3\over 2}
(4 m_u m_s + {7\over 3}\beta_s^2)
\nonumber\\
\hat{\bf N}_4&=&0
\label{ki_amp_ann1}
\end{eqnarray}

Using these equations and
the values of parameters mentioned above we obtain
the following values of $r$ (defined in eqn (\ref{r_def}))
for the contribution of the annihilation graphs to
various channels:
\begin{equation}
\begin{tabular}{rclcrcl}
$r(\gamma      k_0)$&$=$&$5.3\times 10^{-5}$&\ \ \ \ \ \ &
$r(\gamma \hat k_1)$&$=$&$1.1\times 10^{-6}$\\
$r(\gamma \hat k_2)$&$=$&$3.3\times 10^{-5}$&\ \ \ \ \ \ &
$r(\gamma      k_3)$&$=$&$6.6\times 10^{-5}$\\
$r(\gamma      k_4)$&$=$&$0$&&&&
\end{tabular}
\label{isgwresults}
\end{equation}

Our model independent attempt to estimate
the value of $r$ is based on
projecting the component of the quark amplitude with the same quantum
numbers as $k_i$ and then converting these
components to a decay rate as we shall discuss below.

With respect to the spin and angular degrees of freedom in the
meson rest frame we can define the following eigenstates
of angular momentum:
\begin{eqnarray}
|1^-   >&=&Y^0_0    |1\ -1>_s     \nonumber\\
|1^{+-}>&=&Y^{-1}_1 |0\ 0>_s      \nonumber\\
|1^{++}>&=&{1\over\sqrt{2}}( Y^{-1}_1 |1\ 0>_s-Y^{0}_1 |1\ -1>_s)
\nonumber\\
|2^{+} >&=&{1\over\sqrt{2}}( Y^{-1}_1 |1\ 0>_s+Y^{0}_1 |1\ -1>_s).
\end{eqnarray}
where $|s m>_s$ represents the spin state of the $\bar u s$ system.

For each of these states let
$\fanN(|i>)$ be the corresponding amplitude.
We may then construct the quantity
$d r_{|i>}/ds$.
We now estimate the value of
$r(\gamma k_i)$
using
\begin{equation}
r(\gamma k_i)=
\int_{s_{min}}^{s_{max}} {d r_{|i>} \over ds}ds
\end{equation}
for suitably chosen values of $s_{min}$ and $s_{max}$.

For the four states $k_{1,\dots,4}$
we note that the next  $k_i$ states above them is at $1.950 GeV$
with width $.2 GeV$
so it seems reasonable to chose $\sqrt{s_{max}}=1.750 GeV$.
Intuitively
a $\bar u s$ system with invariant mass below
this threshold is forced to form the $k_i$ state with the appropriate
quantum numbers. The results for these states do not depend strongly on
the value of $s_{min}$ so we take $s_{min}=0$.
likewise for $k_0$ we take the threshold $s_{max}=1.2 GeV$ since the
next similar state ($k_3$) has mass about $1.4 GeV$ and width about
$.2GeV$.
Using these thresholds we obtain the following $r$ values:
\begin{equation}
\begin{tabular}{rclcrcl}
$r(\gamma      k_0)$&$=$&$1.2\times 10^{-6}$&\ \ \ \ \ \ &
$r(\gamma \hat k_1)$&$=$&$3.9\times 10^{-7}$\\
$r(\gamma \hat k_2)$&$=$&$3.8\times 10^{-5}$&\ \ \ \ \ \ &
$r(\gamma      k_3)$&$=$&$6.5\times 10^{-5}$\\
$r(\gamma      k_4)$&$=$&$0$&&&&
\end{tabular}
\end{equation}
As is apparent, the
values are similar to those above in
(\ref{isgwresults})
except for the case of $k_0$ which
will not effect our
results greatly
as it is separated too far from the other resonances to
interfere
to any large extent.

\subsection{Spectator Diagram}

Let us now consider the decay rate which is generated by the spectator
graph.
For this purpose we first calculate the quark level process and then
estimate the resultant meson formation.
The quark level reaction at tree level proceeds
through four diagrams
similar to the one
shown in figure 1c.
For the purposes of our calculation
we approximate the
final state quarks as being massless
and so it is convenient
to calculate the helicity amplitudes for the processes.

Let us define a convention for spinors and photon polarizations similar
to those used in \cite{hel_ref}. Our conventions will be based on
three arbitrary light-like reference vectors
$\lambda_0$, $\lambda_1$ and $\lambda_2$. Let $u_0$ be
a right handed spinor in the direction $\lambda_0$ so that

\begin{equation}
u_0\bar u_0=P_R\slam_0.
\end{equation}

\noindent Likewise we define the left handed spinor $u_1$ in direction
$\lambda_1$ as

\begin{equation}
u_1={\slam_1 u_0\over\sqrt{2\lambda_1\cdot\lambda_2}}
\end{equation}

\noindent For a general light-like vector $p$ let us define the left
and right handed spinors $u_-$ and $u_+$ respectively:

\begin{equation}
u_-={\slp u_0\over \sqrt{2 p\cdot \lambda_0}  } \ \ \ \
u_+={\slp u_1\over \sqrt{2 p\cdot \lambda_1}  }
\end{equation}
we will not include the color indices throughout.

For simplicity let us adopt the notation

\begin{equation}
[p_1,\dots,p_n]_\pm =\bar u_\#(p_1) \slp_2 \dots \slp_{n-1} u_\pm(p_n)
\end{equation}

\noindent where $u_\#(p_1)=u_\pm(p_1)$ if $n$ is odd and
$u_\#(p_1)=u_\mp(p_1)$ if $n$ is even. Here $p_1$ and $p_n$ are assumed
to be massless while  $p_2\dots p_{n-1}$ need not be.
Using the definitions of the spinors, we can expand this notation in terms
of traces:

\begin{equation}
[p_1,\dots,p_n]_+= \left\{ \begin{array}{cl}
{Tr(\slp_1\dots\slp_n\slam_1 P_R)
\over\sqrt{4\ \lambda_1\cdot p_1\ \lambda_1\cdot p_n}}
& {\rm if\ {\it n}\ is\ odd} \\
{Tr(\slam_0 \slp_1\dots\slp_n\slam_1 P_R)
\over\sqrt{4\ \lambda_0\cdot p_1\ \lambda_1\cdot p_n}}
& {\rm if\ {\it n}\ is\ even}
\end{array} \right.
\end{equation}

\noindent and the corresponding expression for $[\ \ ]_-$ is obtained
by changing $P_L\leftrightarrow P_R$ and $\lambda_0\leftrightarrow
\lambda_1$.

Circularly polarized photons may also be expressed in this notation. Thus
for a photon with momentum $q$ we may write

\begin{equation}
E_R^\mu={\bar u_+(\lambda_2)\gamma^\mu u_+(q) \over
2\sqrt{\lambda_2\cdot q}}.
\end{equation}

\noindent Left handed polarized photons may be
expressed as $E_L^\mu=E_R^{\mu*}$.

The spinor for a massive fermion with mass $m$ momentum $p$
and spin $s$ can be expanded in terms of massless  spinors as follows:

\begin{eqnarray}
u(p,s)&=&u_-(p_-)+{1\over m}[p_+,p_-]_-u_+(p_+)\nonumber\\
v(p,s)&=&u_-(p_-)-{1\over m}[p_+,p_-]_-u_+(p_+)
\end{eqnarray}

\noindent where $p_\pm=(p\pm m s)/2$

Let us now define the function

\begin{eqnarray}
\fanL(p_1,q,p_2,p_3,p_4,p_5,p_6)&=&\ \nonumber\\
&&[ \bar u_-(p_2)\gamma^\mu \slq \gamma^\nu u_-(p_1)]
\ [ \bar u_+(p_4)\gamma_\nu u_+(p_3)  ]\nonumber\\
&&[ \bar u_-(p_6)\gamma_\mu u_-(p_5)]\nonumber\\
&=&-2{[p_2p_6p_5]_-[p_6p_5qp_3]_+[p_4p_1]_- \over
p_5\cdot p_6}
\end{eqnarray}

We may now write the expression for the $b$ spectator process
$b(p_b)\to u(p_1)\bar u(p_2) s(p_3)\gamma(q)$ for a  left polarized
photon as:
\begin{eqnarray}
\fanM(\gamma_L)&=& {e.g^2_W
V_{us}^*V_{ub}\over 24 m_W^2\sqrt{q\cdot\lambda_2}} \bigopen {2\over
q\cdot p_1}  \fanL^*(p_1,q+p_3,p_{b-},\lambda_2,q,p_3,p_2)\nonumber\\
&&-{1\over  q\cdot p_3} \fanL^*(p_3,q,p_2,\lambda_2,q,p_1,p_{b-})
-{2\over  q\cdot p_2} \fanL (p_2,q,p_3,q,\lambda_2,p_{b-},p_1)\nonumber\\
&&+{1\over  q\cdot p_b} \fanL(p_{b-},p_{b-}-q,p_1,q,\lambda_2,
p_2,p_3)\nonumber\\
&&+{1\over  q\cdot p_b} (p_2\cdot p_3\ \ \lambda_2\cdot q)^{-1} [p_1 p_3
p_2]_-[\lambda_2 q p_{b+} p_{b-}]_-[p_3p_2q]_- \bigclose.
\label{specamp}
\end{eqnarray}
where $g_W$ is the weak coupling constant given in terms of the fermi
coupling $G_F$ by:
\begin{equation}
G_F={g_W^2\over \sqrt{32}m_W^2}
\end{equation}
The amplitude for right polarized photons is given by
interchanging $q\leftrightarrow \lambda_2$.

QCD corrections of course come into play,
the value of $r$ obtained from
(\ref{specamp})
need only be multiplied by a factor of:
\begin{equation}
F={1\over 3}
\left(
2(c_++c_4+c_3)^2+(c_+c_4-c_3)^2
+
2(c_6+c_5)^2+(c_6-c_5)^2
\right)
\end{equation}
from the results in equation (\ref{cresults})
we see that
$F=1.073$ so in fact the tree level description seems
to be reasonable.

Given the amplitudes
for the tree graphs discussed in the last sections,
we must now estimate the formation of various
particular $k_i$ resonances.
We will do this using a ``hand waving'' argument along the lines of our
cutoff method above.
Thus, we will use the following two
simplifying assumptions:
(1) if a system of quarks has the correct
quantum numbers to form a particular resonance and the invariant mass
of the system is within a reasonable range of the resonance (which we
shall define) then it will form the given resonance.
(2) A system of quarks will couple most strongly to the lowest orbital
angular momentum state possible.

Thus, in the case
of the spectator graph we need to distinguish
two separate cases
(a) the $b$ quark has spin $S_z(b)=-{1\over 2}$ and
(b) the $b$ quark has spin $S_z(b)=+{1\over 2}$.
Note that this discussion applies equally to $B^-$ or $\bar B^0$
decays.

In case (a)
the spectator $\bar u$ quark
must be polarized $S_z(\bar u_{spec})=+{1\over 2}$.
Since we are considering only the left handed photon, the spin
projection of the hadronic system
$J_{z}(h)=-1$.
The left handed nature of the coupling implies that
the spins of the quarks from the decay of the $b$ quark
are $S_z(s)=-{1\over 2}$, $S_z(u)=-{1\over 2}$ and
$S_z(\bar u_{part})=+{1\over 2}$.
In total the spin of the quarks
forming the hadron
is $S_z(h)=0$ hence $L_z(h)=-1$.
{}From our assumption (2) it follows that $L(h)=1$
and therefore the $J^P=1^-$.
Thus in this case the preferred final states are $k_0$ or $k_3$.

In case (b) the spectator $\bar u$ must be polarized
$S_z(\bar u_{spec})=-{1\over 2}$ while all the participant
quarks have the same $S_z$ as before.
Thus for a left handed photon,
we have $L_z(h)=0$.
and
from assumption (2)
therefore
$L(h)=0$ so that the total
$J^P=1^+$ and $\hat k_1$ and $\hat k_2$ should be the favored states.

We
can
be
somewhat
more
specific
by
noticing
that
for example if the decay occurs through
$O_2$ the
spectator $\bar u$
quark and the $u$ quark have the same color.
If it should further happen that the remaining pair of
quarks have a different color, then the $u \bar u$ pair
have the quantum numbers of a $\rho$ while the remaining pair have the
quantum numbers of $K$.
On the other hand, only in the situation where the
two $\bar u$'s (the spectator and the one derived from the virtual $W$)
which have opposite spins happened to have the same color could
they form a system which had the quantum numbers as a $K^* \pi$.
In fact the color
part of the amplitude for the other pairing
in the
$\pi K^*$ configuration is
3 times smaller than that of the $K\rho$.
We assume that the production of the final state
$\hat k_1$ and $\hat k_2$
are in proportion to the
coupling to the
$K^*\pi$ or $K\rho$
like  configuration of the
quarks times the coupling of these $1^+$ states to the
$K^*\pi$ or $K\rho$.
The situation for $O_1$ is the same except the $u$ and $s$ quark are
interchanged
as are the
$K^*\pi$ and $K\rho$ final states
and so a fermionic $-$ sign must
also be inserted.

Let us
denote
$\hat b_1^{\rho K}$, $\hat b_1^{\pi K*}$,
$\hat b_2^{\rho K}$ and
$\hat b_2^{\pi K*}$
to be respectively the
coupling of $\rho K$ and $\pi K^*$ to $\hat k_1$ and $\hat k_2$.
{}From the above argument therefore the ratio between
the spectator amplitudes
$\hat \fanM^{spec}_1$ and
$\hat \fanM^{spec}_2$ is
\begin{equation}
\hat \fanM^{spec}_1:
\hat \fanM^{spec}_2 =
c_1 \hat b_1^{\pi K*}-c_2 \hat b_1^{\rho K}:
c_1 \hat b_2^{\pi K*}-c_2 \hat b_2^{\rho K}
\end{equation}

Using the values in equation (\ref{cresults}) and the couplings derived
from experiment derived in
reference \cite{kspec}
the above ratio is
\begin{equation}
\hat \fanM^{spec}_1:
\hat \fanM^{spec}_2 =
14:-1
\end{equation}
so
most of the amplitude is in the $\hat k_1$ channel.

In order to use assumption (1) we need to decide what threshold to use.
Following our discussion above we again pick the threshold
$s_{max}=(1750\mbox{ MeV})^2$.
In the case of  the annihilation graph, the threshold applies to
$s_h=(p_u+p_s)^2$ while in the case of the spectator graph
$s_h=(p_1+p_2+p_3+p_4)^2$ where $p_4$ refers to the momentum of the
spectator $\bar u$. In our formulation of the amplitude of the $b$-quark decay
we can directly calculate the differential cross section in terms of the
variable $s_3=(p_1+p_2+p_3)^2$. If we assume that the spectator quark is
roughly stationary then $p_4=x_u P_B$ and the relation between the
quantities is
\begin{equation}
s_h=(m_b+m_u)(m_bm_u+s_3)m_b^{-1}
\end{equation}
thus $s_h\leq (1750\mbox{ MeV})^2$ translates to $s_3\leq
1120$ MeV\null.

Therefore the
values of $r$ for  the spectator graph is
$5\times 10^{-7}$  for  case  (a) where $S_z(b)=-{1\over 2}$ and
$1.1\times 10^{-5}$ for  case  (b) where $S_z(b)=+{1\over 2}$.
Thus, from
the spectator graph we
obtain the following values:
\begin{equation}
\begin{tabular}{rclcrcl}
$r(\gamma      k_0)$&$=$&$5  \times 10^{-7}  $&\ \ \ \ \ \ &
$r(\gamma \hat k_1)$&$=$&$1.1\times 10^{-5}$\\
$r(\gamma \hat k_2)$&$=$&$5.6\times 10^{-8}$&\ \ \ \ \ \ &
$r(\gamma      k_3)$&$=$&$5  \times 10^{-7}$\\
$r(\gamma      k_4)$&$=$&$0                $&&&&
\end{tabular}
\end{equation}

\subsection{Meson Couplings}

In order to guide our calculations let us now estimate the CP phase
and couplings for each of the five channels which we consider.
Let us denote $\fanB_{pen}$ to be the inclusive
branching ratio of $b\to s\gamma$ through the penguin graph and
$\fanB_{bus}$
to be $\Gamma_{bus}^0\over\Gamma_B$.
In our numerical calculations, for concreteness we will take,
$\fanB_{pen} = 2.5\times 10^{-4}$
roughly corresponding to $m_t=150GeV$
\cite{cleobsg}.
If we take
$|{V_{ub}\over V_{cb}}|=.08$
\cite{cleobu}, $V_{us}=.22$
and the leptonic branching ratio to be
0.107
then from equation
(\ref{gam0def})
we obtain
$\fanB_{bus}=1\times 10^{-4}$.
(Of course, the rate for the process $b\to u\bar su\gamma$ will be
significantly smaller than this)

For a given channel $k_i$ which
decays to a final state $XY$ we model the
contribution to the decay process

\begin{equation}
B\to k_i\gamma \to XY\gamma \label{react}
\end{equation}

\noindent in two stages. Thus we define a
coupling $A_i$ governing the decay
$B\to k_i\gamma$ and the coupling $b_i$ governing  $k_i\to XY$.
The amplitude for the entire process (\ref{react}) is thus
$A_i\Pi_{ij}b_j$ where $\Pi_{ij}$ is the propagator to be discussed later.

Our model for this process will be such that
all the interaction phase is in $\Pi_i$
while all the CP phase is in $A_i$.
Thus $A_i$ is a complex coupling
which we may express as:

\begin{equation}
A_i=a_i e^{i\phi_i}
\end{equation}

\noindent where $\phi_i$ is the CP phase and therefore flips sign under
charge conjugation.
Note that $\phi_i$ is related to the CKM phase parameter $\delta$
as
explained below.

{}From the decay rates which we have estimated for the penguin and tree
processes separately, we may determine the total amplitude, $A_i$:
\begin{eqnarray}
A_i&=&
\sigma_p^i \sqrt{ 16\pi m_B^3\fanB_{pen}R_i\Gamma_B\over (m_B^2-m_i^2)}
+
\sigma_{ann}^i \sqrt{ 16\pi m_B^3\fanB_{bus}r^{ann}_i\Gamma_B\over
(m_B^2-m_i^2)}
e^{i\delta}\nonumber\\
&&+
\sigma_{spec}^i \sqrt{ 16\pi m_B^3\fanB_{bus}r^{spec}_i\Gamma_B\over
(m_B^2-m_i^2)}
e^{i\delta}
\label{Adef}
\end{eqnarray}
where $R$ is defined in equations (\ref{R_def}), $\delta$ is the
CP phase defined in equation (\ref{vkmdef})
and $r^{ann}_i$ and $r^{spec}_i$ are the values from the annihilation and
spectator graphs obtained above.
$\sigma_p^i$, $\sigma_{ann}^i$
and
$\sigma_{spec}^i$
are the relative signs between the three amplitudes and are thus
either $\pm 1$.
In the case of
$\sigma_p^i$
equation \ref{penampli}
gives $\sigma_p^i=-1$ for each of the
states; model B gives similar results.
Equation
\ref{ki_amp_ann1} gives $\sigma_{ann}^i=-1$ for
$i\in\{1,2\}$
and  $+1$ for
$i\in\{0,3 \}$; the projection method gives the same results.
In our model
$\sigma_{spec}^i$
is undetermined though
it has very little effect on our final results
as the spectator amplitude is too small compared to
the annihilation or the penguin amplitudes.
In our numerical calculations, we will assume
in the first instance, that
$\sigma_{spec}^i=\sigma_{ann}^i$.
Later, in section 3.4, we will try to determine these
relative signs between the amplitudes by using a simple model.
Furthermore, we will also numerically investigate
the effect of switching the signs.
Note that
the penguin graph is the dominant
production mechanism for $B\to k_i\gamma$,
hence the magnitude $a_i$ is given to a good approximation by
the first term in (\ref{Adef})

As we mentioned before
the CP phase,
which in our convention is $\delta$,
is contained in the tree processes.
We can therefore estimate the total phase $\phi_i$ by:

\begin{equation}
\left | {\sin\phi_i\over\sin\delta} \right | =\sqrt{\fanB_{bus}
r_i\over \fanB_{pen} R_i}
\end{equation}

\noindent where $r_i$ is defined in equation (\ref{r_def}).
The
numerical results for the cases which we consider are compiled in Table 3.

Next we deal with the couplings of the strong
decays of the resonances leading to the final states.
Their couplings $b_i$
for $i=0$, $3$ and $4$
may be obtained from the meson decay
widths:
\begin{equation}
b_i=\sigma_d^i \sqrt{16\pi m_i^3 Br(k_i\to XY)
\Gamma_i\over \lambda^{1\over
2}(m_i^2,m_X^2,m_Y^2)}
\end{equation}
Here $\Gamma_i$ is the total width of $k_i$ and
\begin{equation}
\lambda(u,v,w)=u^2+v^2+w^2-2uv-2vw-2wu. \nonumber
\end{equation}
Again
$\sigma_d^i$
is $\pm 1$.
In the next section we will discuss in more detail how
this sign may be determined.

In \cite{kspec} the couplings to the physical states $k_1$ and $k_2$
are parameterized in terms of $\thx$, $\gamma_+$ and $\gamma_-$:
\begin{eqnarray}
b_1^{K*\pi}&=& -{1\over 2}\gamma_+\sin\thx
+\sqrt{9\over 20}\gamma_-\cos\thx \nonumber\\
b_2^{K*\pi}&=& +{1\over 2}\gamma_+\cos\thx
+\sqrt{9\over 20}\gamma_-\sin\thx \nonumber\\
b_1^{K\rho}&=& -{1\over 2}\gamma_+\sin\thx
-\sqrt{9\over 20}\gamma_-\cos\thx \nonumber\\
b_2^{K\rho}&=& +{1\over 2}\gamma_+\cos\thx
-\sqrt{9\over 20}\gamma_-\sin\thx
\label{couplings12}
\end{eqnarray}

where the observed values of these $\gamma's$ are\cite{kspec}:
\begin{equation}
\gamma_+=0.82\ \ \ \
\gamma_-=0.59\ \ \ \
\thx=56^\circ
\end{equation}

\eject

\section{How CP Violation May be Detected}

Let us consider the various  two body decay modes of the $k_i$ states
listed in  Table 1. Our basic strategy will be to consider the process
$B\to \gamma k_i\to\gamma XY$ where more than one possible
intermediate state $k_i$ may occur. If the different states have
different quantum numbers although the final states are the same, the
angular distributions will be different. In addition if there are
different interaction phases and CP phases associated with each $k_i$
state there will be a difference in the angular distribution  between
the decay products of $B$ and $\bar B$ which therefore signal CP violation.
Indeed if the  quantum
numbers are the same, as in the case of $k_0$ vs.\ $k_3$ and $k_1$ vs.\
$k_2$, the interference could lead to a partial rate asymmetry.

For such two body decays,
let us define $s_h$ to be the invariant mass of the $k_i$ state,
$s_h=(P_X+P_Y)^2$ and let $\theta$ be the angle between the boost axis
and the momentum of the strange particle
($K$ or $K^*$)
in the $k_i$ rest frame and let $\phi$ be the azimuthal angle.
Denoting
\begin{equation}
z=\cos\theta
\label{z_def}
\end{equation}
the energy of $X$ in the $B$ rest frame, is
given b:

\begin{equation}
E_X= {(m_B^2+s_h)(s_h+m_X^2-m_Y^2)
-(-1)^{{\bf s}_X}
z(m_B^2-s_h)\lambda^{1\over 2}(s_h,m_X^2,m_Y^2) \over 4m_Bs_h}
\end{equation}
where ${\bf s}_X$ is the strangeness of $X$.
Using these variables we denote the decay distributions

\begin{equation}
G(\sh,z)={d^2\over d\sh dz}\Gamma(B\to\gamma XY)\ \ \ \
\bar G(\sh,z)={d^2\over d\sh dz} \bar \Gamma(B\to\gamma XY)
\end{equation}

\noindent Of interest to us are the sum and the difference of these
quantities:

\begin{equation}
\Delta(\sh,z)=G(\sh,z)-\bar G(\sh,z)\ \ \ \
\Sigma(\sh,z)=G(\sh,z)+\bar G(\sh,z) \label{sigdel_ref}
\end{equation}
A non-zero value of $\Delta $ is clearly CP violating.

Examining Table 1 it is evident that there are many cases where different
channels can lead to the same final state and therefore
CP violating energy distributions may be possible.
In particular there are two classes of final states which we will consider;
$k_i\to XY$ where $X$ is a vector and $Y$
is a pseudoscalar
(ie either $\rho K$ or $\pi K^*$)
and the case
$k_i\to UV$ where both $U$ and $V$ are pseudoscalars
(ie. $K\pi$).

\subsection{Interference Between $k_i$ Resonances}

All of the observables which we consider here are based on the
distribution $\Delta$ being non-zero. In the case of the interference
between the two $k$ states with different quantum numbers,
$\Delta$ will arise due to the diagram in Figure 2a. The blobs in this
diagram indicate the rescattering which
produces an imaginary part of the propagator.

In the case of the interference between $k_i$ and $k_j$ where these two
states have the same quantum numbers, there is the additional
possible graph in Figure 2b where one state rescatters to the other.
Indeed these graphs are very important since the CPT theorem implies
that the total decay rate of $B$ and $\bar B$
must be the same \cite{pra_ref}.
Hence for a particular final state $f$ which has a
partial rate asymmetry
(PRA) then there
must be some other final state $g$ which has a compensating PRA\null.
To see how this is implemented in the two diagrams consider,
for example, the final state
being $f$ in figure 2a. A contribution to the PRA of $f$
arises from the rescattering of state $k_i$ through state $g$.
This is related to the contribution of figure 2b
to the PRA of $g$
where $k_j$ rescatters to $k_i$ through state $f$. In fact
these two can be shown toø 
be opposite through the Cutkosky relations
and thus will exactly cancel.

In order to understand this properly let us consider the instance of
two interfering states $k_i$ and $k_j$ giving rise to the PRA of
state $f_l$.
We can break down the amplitude in this instance into three parts.
The decay $B\to \gamma k_i$, the propagation of $k_i$ and  the decay of
$k_i\to f_l$.

The decay $B\to \gamma k_i$, $\gamma k_j$ may be described by the
amplitudes

\begin{equation}
A=\left(  \begin{array}{c} A_i\\ A_j \end{array} \right).
\end{equation}

\noindent which may contain CP phases.
We can  write the propagator for the two $k$ states as a matrix

\begin{equation}
\Pi= \left( \begin{array}{cc} \Pi_{ii}&\Pi_{ij}\\
\Pi_{ji}&\Pi_{jj}  \end{array} \right)
\end{equation}

\noindent and we represent the
strong interaction decay of the $k$ states by  the amplitudes
\begin{equation}
b^l=\left(  \begin{array}{c} b_{i}^l \\ b_{j}^l \end{array} \right).
\end{equation}
which are real since there is no CP
phase in this instance and we
assume that the absorbtive phase is
contained in $\Pi$.

The amplitudes for $B$ and $\bar B$ decays are thus:

\begin{equation}
\fanM_l= A^T \Pi b^l;\ \ \ \  \bar\fanM_l= A^\dagger\Pi b^l
\end{equation}

\noindent $\Delta$ is thus related to

\begin{eqnarray}
\Delta\fanM_l^2&=&|\fanM_l|^2-|\bar\fanM_l|^2 \nonumber\\
&=&{\rm Tr}\left(  (A^* A^T-A A^\dagger) \Pi b_l b^T_l \Pi^\dagger
\right). \label{resolx}
\end{eqnarray}

Let us consider now what the structure of $\Pi$ is. For the optical
theorem to be true for any possible initial state, $\Pi$ must satisfy
the Cutkosky relation:

\begin{equation}
-Im(\Pi)=\Pi \epsilon \Pi^\dagger \label{cuta}
\end{equation}

\noindent where for the matrix $\Pi$, $Im(\Pi)={1\over 2i}
(\Pi-\Pi^\dagger)$ and
$\epsilon$ is the rescattering matrix defined by:

\begin{equation}
\epsilon_{st}= \sum_l \int  b_{s}^l b_{t}^{l\dagger} d\phi_l .
\label{epsdef}
\end{equation}

\noindent Here the sum is over all possible final states and
the integral is over
the appropriate phase space $\phi_l$ for the final state $l$.

We can thus rearrange equation (\ref{cuta}) to

\begin{equation}
Im(\Pi^{-1})=\epsilon
\end{equation}

Note that $\epsilon$ is real since b is real. Since $\epsilon$ is also
hermitian, therefore it is symmetric too, and so is $Im(\Pi^{-1})$.
\noindent Let us write

\begin{equation}
Re(\Pi^{-1})=s_h-M
\end{equation}

\noindent where $M$ is a mass matrix. We choose
the basis of
the $k$ states so that $M$ is diagonal. Thus

\begin{equation}
M=\left( \begin{array}{cc} m_i^2 & 0\\
0& m_j^2 \end{array} \right)
\end{equation}

and consequently $Re(\Pi^{-1})$ is also symmetric.
Since both
$Im(\Pi^{-1})$
and
$Re(\Pi^{-1})$
are symmetric, so is $\Pi$.

Returning now to equation (\ref{resolx}) we can verify the demand
of CPT that the sum of all PRA's must vanish.
To see this
we note that if we sum over all final states $l$ and integrate over the
phase space of the final state the second factor becomes,
after application of equation (\ref{cuta}), $Im(\Pi)$. Thus
\begin{equation}
\Delta \fanM^2_l = {\rm Tr}\left[ \left( (A^*A^T)-(A^*A^T)^T \right)
{Im\Pi} \right].   \label{resoly}
\end{equation}

\noindent which vanishes as $Im\Pi$
is symmetric whereas its coefficient is anti-symmetric.
The requirement of CPT is therefore confirmed.

If we apply this formalism to the more general case where
the $k_i$ states of distinct quantum numbers are present, then
we may also include components of the distribution $\Delta$ which
do not contribute to the PRA but nonetheless are CP violating.

Furthermore, if a particular state
is the only one with a given set of quantum numbers
contributing to the final state the above formalism
gives the standard Breit-Wigner form:

\begin{equation}
\Pi_i={1\over s-m_i^2 + i\Gamma_i m_i}
\end{equation}

Note that
with respect to the $1^+$ states
in equation (\ref{resolx})
we have worked in the mass basis
$\{ k_1, k_2\}$.
The calculation of the production is
however most naturally carried out in the
quark model basis
$\{ \hat k_1, \hat k_2\}$.
If we denote
$\hat A$ the production amplitude
in the quark model basis
and $\fanF $ the suitable mixing matrix
then we can relate $A$ to $\hat A$ by $A=\fanF\hat A$ in
equation (\ref{resolx}).

\subsection{ Vector Pseudoscalar Case}

First let us consider the case where $k_i\to XY$; $X$ is a vector and
$Y$ is a pseudoscalar. From Table 1 we see that this
happens for the resonances
$\{k_1,k_2,k_3,k_4\}$.
Consider first the case of $1^+$. The quantum numbers dictate that the
decay may proceed through $L=0$ or $L=2$.  Recalling that the $k_i$ has
$J_z=-1$
(as mentioned before the $z$ axis is antiparallel to $\vec p_\gamma$),
thus the decay distribution is proportional to

\begin{equation}
Y_0^0 X_{-1}
\end{equation}

\noindent for the $L=0$ channel where $X_i$ means vector $X$ with
polarization $i$ and $Y^i_j(\theta,\phi)$ is the spherical harmonic.
For the $L=2$ channel the corresponding amplitude is

\begin{equation}
\sqrt{1\over 10} Y^0_2    X_{-1} -\sqrt{3\over 10} Y^{-1}_2 X_{0}
+\sqrt{3\over  5} Y^{-2}_2 X_{+1}
\end{equation}

\noindent In the case of a $2^+$ channel,
$L=2$ and so the decay distribution is  proportional to

\begin{equation}
\sqrt{1\over   2} Y^0_2    X_{-1} -\sqrt{1\over  6}
Y^{-1}_2 X_{0} -\sqrt{1\over  3} Y^{-2}_2 X_{+1} .
\end{equation}

\noindent Finally, in the case of a $1^-$ channel, $L=1$ and hence the
decay distribution is proportional to

\begin{equation}
\sqrt{1\over 2} \left ( Y^0_1   X_{-1} -Y^{-1}_1 X_{0} \right ).
\end{equation}

\noindent Expanding these amplitudes in terms of $z$ and $\phi$ we get

\begin{eqnarray}
\fanM_1&=& b_1\  \sqrt{1\over 4 \pi} X_{-1} \nonumber\\
\fanM_2&=& b_2\  \sqrt{1\over 4 \pi} X_{-1} \nonumber\\
\fanM_3&=& b_3\  \sqrt{3\over 8 \pi} \paropen z
   X_{-1}-\sqrt{1\over 2}\sqrt{1-z^2}e^{-i\phi}X_0 \parclose \nonumber\\
\fanM_4&=& b_4\  \sqrt{5\over 32 \pi} \paropen (3z^2-1) X_{-1}
   \nonumber\\
&&-\sqrt{2}z\sqrt{1-z^2}e^{-i\phi} X_{0} -(1-z^2)e^{-2i\phi}
X_{+1} \parclose
\end{eqnarray}

\noindent where we define the couplings between $k_i$ and $XY$ to be
$b_i$. Thus the matrix $U$ defined above is given by  $U_{ij}=b_i b_j
\fanR_{ij}$ where

\begin{equation}
\fanR_{(XY)}=\left( \begin{array}{ccccc}
0&0&0&0&0\\
0&{1\over 2} & {1\over 2}&\sqrt{3\over 8}\ z& \sqrt{5\over 32} (3z^2-1) \\
0&{1\over 2}& {1\over 2}& \sqrt{3\over 8}\ z& \sqrt{5\over 32} (3z^2-1) \\
0&\sqrt{3\over 8}\ z & \sqrt{3\over 8}\ z &
  {3\over 8}(1+z^2)& \sqrt{15\over 16} z^3\\
0&\sqrt{5\over 32} (3z^2-1) &\sqrt{5\over 32} (3z^2-1) &
\sqrt{15\over 16} z^3& {5\over 8}(4z^4-3z^2+1)
\end{array} \right)
\end{equation}

\subsection{Pseudoscalar-Pseudoscalar case}

Now let us consider the case where $k_i\to UV$ and $U$, $V$ are
pseudoscalars;
in particular,  $\pi$, $K$.
In this instance the only states
involved are $k_0$, $k_3$ and $k_4$. The
relevant amplitudes are as follows:

\begin{eqnarray}
\fanM_0 &=& b_0\sqrt{3\over 8 \pi}\sqrt{1-z^2}e^{-i\phi} \nonumber\\
\fanM_3 &=&b_3\sqrt{3\over 8 \pi}\sqrt{1-z^2}e^{-i\phi} \nonumber\\
\fanM_4 &=& b_4\sqrt{15\over 8 \pi}z\sqrt{1-z^2}e^{-i\phi}
\end{eqnarray}

\noindent Hence the corresponding matrix $\fanR$ is given by:

\begin{equation}
\fanR_{(UV)}= \left( \begin{array}{ccccc}
{3\over 4}(1-z^2)&0&0&{3\over 4}(1-z^2) & \sqrt{45\over 16}z(1-z^2)\\
0&0&0&0&0\\
0&0&0&0&0 \\
{3\over 4}(1-z^2)&0&0&{3\over 4}(1-z^2)& \sqrt{45\over 16}z(1-z^2)\\
\sqrt{45\over 16}z(1-z^2)&0&0& \sqrt{45\over 16}z(1-z^2)& {15\over
4}z^2(1-z^2) \end{array} \right)
\end{equation}

\subsection{Signs of Decay Amplitudes}

The convention which we used for the angular variables $\theta$
is constructed such that
if all the amplitudes are positive then constructive interference occurs
if the final strange meson ($K$ or $K^*$)
is in the forward ($+z$) direction.
Bearing this convention in mind,
let us consider a crude model, based loosely on the idea
of ``vacuum dominance'', which we will use,
for the sole
purpose of suggesting the signs of the decay amplitudes.
As an illustration, let us consider the $K^*\pi$ final state. The full reaction
is thus:
\begin{equation}
B\rightarrow \gamma k_i\rightarrow \gamma \pi K^*
\end{equation}
The contributing $k_i$ are then $k_0$, $k_3$ and $k_4$ (see Table 1).
Let us concentrate on the case when the $B \rightarrow \gamma k_i$
decay takes place via the penguin graph. Then the vacuum saturation
representation
is
\begin{equation}
\fanA = Tr\left( \Pi_b \gamma_5 \Pi_{u1} \Gamma_i \Pi_{s1}
\sigma^{\mu\nu} P_R\right)
Tr\left(\Pi_{s2}\Gamma_i \Pi_{u2} \gamma_5 \Pi_d \slee\right)
\label{avd}
\end{equation}
where
$\Pi_b$, $\Pi_{u1}$ and $\Pi_{s1}$ are propagators of the
$b$, $\bar u$ and $s$ quarks in the $B\rightarrow k_i\gamma$ decay;
$\Pi_d$, $\Pi_{u2}$ and $\Pi_{s2}$ are propagators of the
$d$, $\bar u$ and $s$ quarks in the
$ k_i\rightarrow K^*\pi$ decay and $\Gamma_i$ is the appropriate
gamma matrix insertion for the state $k_i$ and $\slee$ is the polarization
of the $K^*$.

For $i=0$, $3$ we take $\Gamma_i=\gamma^\mu E_\mu$
while for $i=\hat 2$ we take $\Gamma_i=\gamma^\mu \gamma_5 E_\mu$.
In the case of $i=\hat 1$ the relative sign with $i=\hat 2$
is determined from \cite{kspec} as described in equation
(\ref{couplings12}). For the spin $2^+$ case $i=4$ we take
$\Gamma_i=\gamma^\mu E_{\mu\nu} \vec P^\nu$ where
$E_{\mu\nu}$ is the spin 2 polarization
tensor and $\vec P$ is the
momentum of the $s$-quark in the $k_4$ frame.

Let us now consider the configuration where $\theta=0$ and ${\cal E}$
is left handed. Further let us take $p_b= x_b p_B$ and
$p_{s2}=x_s p_{K*}$ where
$x_b=m_b/(m_b+m_u)$ and
$x_s=m_s/(m_s+m_u)$ so that all the other quark momenta are determined by
momentum conservation. We thus find that the signs in this model
are given by:
$\sigma_{d0}=\hat\sigma_{d2}=\sigma_{d3}=\sigma_{d4}=+1$.
Applying a similar analysis to the $\rho K$ final state we find the
same signs hold:
$\sigma_{d0}=\hat\sigma_{d2}=\sigma_{d3}=\sigma_{d4}=+1$
as well as  for the $K\pi$
final state
$\sigma_{d0}=\sigma_{d3}=\sigma_{d4}=+1$.
Although, in our numerical work, for definiteness
we will use signs as given by this simple model,
later we will comment on possible
effects due to the signs being different from those
given by this model.

\subsection{Observables}

In order to observe a component of the asymmetry, it is useful to form
an observable with the same
symmetry as the component we wish to observe.
Thus we can take any function $w_i(s_h,z)$ and form the quantity

\begin{equation}
<w_i>_{B}-<w_i>_{\bar B}
\end{equation}

\noindent which is CP violating. The effectiveness
of this observable to statistically extract the signal from the
background can be parameterized by the quantity:

\begin{eqnarray}
\fanE_i= { \int w_i(s_h,z) \Delta d\sh dz \over \sqrt{
\int w_i^2 \Sigma d\sh dz\  \int\Sigma d\sh dz}}
\end{eqnarray}

\noindent where $\Sigma$ and $\Delta$ are defined in equation
(\ref{sigdel_ref}) and $z$ is defined in equation (\ref{z_def}).

The meaning of this quantity is that  given $N$
events of the specified form
the effect may be distinguished with a significance of
$\fanS=\fanE\sqrt{N}$.
Thus the total number $N_B$ of $B$ mesons (including both $B$ and
$\bar B$) needed to observe the effect at $1-\sigma$
is
\begin{equation}
N_B={1\over {\rm Br}\ \fanE^2}
\label{nb_def}
\end{equation}
where ${\rm Br}$ is the total branching ratio for (\ref{react}).
Clearly we would like $\fanE$ to be as large as possible.
In fact the function which maximizes
$\fanE$
is
\cite{opt_ref}
\begin{equation}
w_{opt}={\Delta \over \Sigma}
\label{optdef}
\end{equation}
Using this observable the expression for $\fanE$ simplifies to

\begin{equation}
\fanE_{opt}=\left({  \int {\Delta^2\over\Sigma}\ ds_h\ dz
\over\int\Sigma\ ds_h\ dz}\right)^{1\over 2}.
\end{equation}

Another  form of observables that we consider are asymmetries where
$w=\pm 1$ at all points in phase space.
In this case the definition of $\fanE$ simplifies to:

\begin{eqnarray}
\fanE= { \int w(s_h,z) \Delta d\sh dz \over \int\Sigma d\sh dz }
\end{eqnarray}

\noindent corresponding to the usual definition of an asymmetry.

Let us now consider three specific asymmetries:

\begin{equation}
w_0=1\ \ \ \ w_1={\rm sign}(z)\ \ \ \ w_2={\rm sign}(|z|-{1\over 2})
\end{equation}

\noindent In Figure 3 we plot the differential asymmetries
$d\fanE_i/d\sqrt{s_h}$ together with $d\Sigma/d\sqrt{s_h}$.
Note that $i=0$ corresponds to $PRA$ and arises
when resonant states with identical quantum contribute to
the same final state; $i=1$ and $i=2$ correspond
to asymmetries in the energy distributions. These
arise when contributing resonance states have the opposite parity
or have the same parity, respectively.

Finally, we note that, in order to enhance the asymmetry observed it
may also be useful to
modify the above as follows:

\begin{equation}
w_i^\prime={\rm sign}(d\fanE_i/ds_h) w_i(z)
\end{equation}

\noindent thus flipping the sign according to the expected sign changes
as a function of $s_h$. However, in the specific cases that
we consider asymmetries do not seem to switch signs as $s_h$
changes so that this sort of multiplication by the sign
turns out not to be useful.

It is instructive at this point to consider how
$\fanE$ and $N_B$ scale with $B_{pen}$.
Consider changing $B_{pen}\rightarrow \lambda B_{pen}$.
In the above formalism then
$\Sigma\rightarrow\lambda\Sigma$ and
$\Delta\rightarrow\lambda^{1\over 2}\Delta$ thus
$\fanE\rightarrow\lambda^{-{1\over 2}}\fanE$
but since ${\rm Br}\rightarrow \lambda {\rm Br}$
$N_B\rightarrow N_B$. Thus $N_B$ is independent of the
exact normalization of the penguin rates.
Furthermore this implies
that $N_B$ is
also relatively independent of the efficiency of forming
$k_i$ states from the penguin process.

\subsection{Numerical Results}

For the purposes of numerical results we take the CKM phase
$\delta$ to be $\pi\over 2$ and
the signs of the amplitudes as described in section 2.5 and 3.4.

For resonance formation from the annihilation graph we use the
potential model results given
in (\ref{isgwresults}).
Our key results are shown in Table 4 and Figure 3.
In Table 4 we use this method to calculate $\fanE_i$ for the above
observables as well as the optimal observable give by equation
(\ref{optdef}) for each of the final states.
By using equation (\ref{nb_def}) and the corresponding branching
ratios,
we also calculate the
number of $B$'s necessary to observe a statistical effect
at the 1 sigma level.
The resulting numbers are also
shown in Table 4; note that these are given in units of $10^8$.

Figure 3a
shows ${1\over \Sigma} {d\Sigma\over d \sqrt{s_h}}$
as a function of $\sqrt{s_h}$. Here the solid line represents
the decay to the $\rho K$ final state, the dotted line to the $K^*\pi$
final state and the dashed line for the $K\pi$ final state.
Note that the peaks correspond to the resonances in Table 1 which are
indicated in the graph by the bars.
In Figure 3b
${1\over \Sigma} {d\fanE_0\over d \sqrt{s_h}}$
is shown. In the $\rho K$ and $K^*\pi$ modes the effects are due
to the interference of $k_1$ and $k_2$.
The resultant values of $\fanE_0$ are also shown in Table 4.
Note
that since the resonances $k_0$ and $k_3$ are so far apart the value of
$\fanE_0$ is negligibly small for the $K\pi$ final state.
Likewise Figure 3c shows
${1\over \Sigma} {d\fanE_1\over d \sqrt{s_h}}$
for each of the final states. For the $K^*\pi$ and $K\rho$ final states,
there is a complex structure since contributions result from the
interference of any pair of resonances with opposite parity.
This also accounts for the large values in $\fanE_1$.
Similar comments are valid for the $K\pi$ state except the $k_1$ and
$k_2$ states are not involved while the $k_0$ is.
With the resonances that we consider the $K\pi$ final state does not
contribute to $\fanE_2$. For the other final states, the curves
in Figure 3d are due to the interference of various positive parity
states with each other.
The optimal $\fanE_{opt}$
and the corresponding values of $N_{opt}$ given in Table 4
show that
these
effects may be
seen with about  $10^9$ $B$'s .

It is apparent that there is considerable uncertainty in these
results; some of which should be reduced in the future. First of all consider
the ratio $R(k_i\gamma)$.
The theoretical prediction based, at the moment, on potential models,
are rather unreliable as our calculations show.
However, this source of uncertainty will get substantially
under control as experimentally samples of about $10^7$ B's
(i.e. well before the $10^8$ or $10^9$ needed for
CP studies )
become available, as then the rates for different resonant
channels will be experimentally measured.
So by the time $10^8$ B mesons have been accumulated
one might anticipate that
many of these ratios will be well determined. Furthermore,
detail fits to the CP-conserving distributions,
e.g.${1\over \Sigma} {d\Sigma\over d s_h}$
as a function of $\sqrt{s_h}$ (see Fig. 3a), of the data that
becomes
available at that time should also allow
the determination of the ambiguities
in the signs of the amplitudes as well
as a more careful determination of the
strong phase than the approximations considered here.

It is important to note that if the penguin
graphs are rescaled by a constant
amount the resultant value of $N_B$,
needed for the CP asymmetry to be observed, is
unaffected. To see this suppose that the
amplitude for the penguin is multiplied  by a factor of $\lambda$.
Since the penguin dominates the production,
the total branching ratio to a
final state varies like $\lambda^2$.
The asymmetry due to interference with the
tree graphs will therefore vary like $\lambda^{-1}$; following equation
(\ref{nb_def}) $N_B$ will therefore be independent of $\lambda$.

Although we have attempted to determine the
signs of the decay amplitude,
it is instructive to consider
the uncertainty in $N_B$ introduced
for other possible sign combinations.
Consider $N_{opt}$ in model A for
the $K^* \pi$ final state. The value with the sign choice indicated above
is
$2.2\times 10^8$ as given in Table 4. If however one checks all the
possible sign combinations one finds that this quantity varies between
$1.5\times 10^8$ and $4.0\times 10^8$.
So once again, our estimate for the required number of B's
is not greatly effected from this source of uncertainty
either.

\section{Conclusions and Summary}

Despite the fact that there are appreciable
uncertainties in our estimates it seems
likely that the asymmetries considered here may be observable in
the case of  $B^\pm$ at a $B$ factory capable of producing
about $10^9$ $B$ mesons.
In our estimates we do not consider what happens for
$s_h$ well above the $k_i$ states. Note especially that the tree graphs
tend to increase rapidly with $s_h$
so that larger CP phases may be available,
though the strong
rescattering phases
and the
branching ratios are likely to become somewhat smaller.
Thus our estimates may well  be underestimates.
Of
course
CP violation may also arise from physics beyond the standard model, and
then too larger asymmetries are possible.

Another point we wish to emphasize briefly has to do with
the formation of the higher resonances in radiative B transitions.
In both bound state models that we studied, we found (as shown in
Table 2) that, except for the $k_1$, the other three states are produced
roughly with the same branching ratio as the $k_0$ (i.e. $K^*(892)$)
which was recently seen experimentally\cite{ref_cleo}.  Experimental
searches of all of these states are vitally needed.

It should be clear that the essential idea proposed in this work is
that resonances can have interesting
and, perhaps, even a dramatic influence on the CP violating
observables. In the case of neutral $B's$ leading to
self-conjugate final states, effects arising from interference
between the initial $B$ and $\bar{B}$ are well known\cite{sanda}.
What is being demonstrated here is that charged B meson
decays leading to common final states via resonances
can also lead to important interference effects. Indeed,
since the underlying theory\cite{km} involves quarks (not mesons)
it is difficult to confine CP
violation just to neutral or just to charged mesons;
resonance
enhancement through such considerations should be possible both
for charged as well as neutral B's.
For concreteness we have, in this paper, only addressed to
the radiative decays of charged $B$'s. Clearly similar effects on neutral $B$'s
need to be
investigated.
Furthermore effects of non-standard physics needs to be ascertained.
We will return to some of these issues in subsequent publications
\cite{atw_soni_prog,japan_talk}.

\bigskip
{\large\bf Acknowledgements}

We are very grateful to Gad Eilam for collaborating with us on the
initial stages of this work.
We also would like to acknowledge useful discussions with
Michael Peskin and
William Dunwoodie. The work of D.~A. is supported by an SSC Fellowship
and DOE contract DE-AC03-76SF00515 while the work of A.~S. is supported
by DOE contract DE-AC02-76CH0016.

\eject

\eject

\begin{center}
{\large\bf Figure Captions}
\end{center}
\bigskip\bigskip\bigskip

\begin{center}
{\bf Figure 1}
\end{center}

\begin{itemize}

\item[Figure 1a:] Example of a penguin graph for the  subprocess for
$b\to s\gamma$.

\item[Figure 1b:] Example of  the annihilation subprocess
$b \bar u \to s\gamma  \bar u$.

\item[Figure 1c:] Example of the spectator subprocess
$b\to s\gamma u \bar u$.

\end{itemize}

\bigskip\bigskip
\bigskip\bigskip

\begin{center}
{\bf Figure 2}
\end{center}

\begin{itemize}

\item[Figure 2a:] A typical instance of two diagrams
contributing to the partial rate
asymmetry of a state $f$. The intermediate state $g$ is shown giving a
contribution to the imaginary part of the propagator.

\item[Figure 2b:] The two diagrams which give the compensating
partial rate asymmetry to the final state $g$ so that CPT is preserved.

\end{itemize}

\bigskip\bigskip
\bigskip\bigskip

\begin{center}
{\bf Figure 3}
\end{center}

\begin{itemize}

\item[Figure 3a:]
A plot of ${1\over \Sigma} {d\Sigma\over d \sqrt{s_h}}$ as a function of
$\sqrt{s_h}$ for the $\rho K $ state (solid line);
$\pi K^*$ state (dotted line) and
$\pi K  $ state (dashed line). The bars indicate the positions of the
five resonances considered.

\item[Figure 3b:]
A plot of ${1\over \Sigma} {d\fanE_0\over d \sqrt{s_h}}$ as a function of
$\sqrt{s_h}$ for the same three final states.

\item[Figure 3c:]
A plot of ${1\over \Sigma} {d\fanE_1\over d \sqrt{s_h}}$ as a function of
$\sqrt{s_h}$ for the same three final states.

\item[Figure 3d:]
A plot of ${1\over \Sigma} {d\fanE_2\over d \sqrt{s_h}}$ as a function of
$\sqrt{s_h}$ for the  $\rho K$ and $\pi K^*$ final states.

\end{itemize}

\eject

\bigskip\bigskip \bigskip\bigskip
\centerline{\bf Table 1}
\bigskip

$$
\begin{tabular}{||lcl|l|l|l|rl||}
\hline
\multicolumn{3}{||c|}{State}           &
\multicolumn{1}{c|}{Mass (Mev)}       &
\multicolumn{1}{c|}{Width (Mev)}      &
\multicolumn{1}{c|}{$^{2S+1}L_J$}     &
\multicolumn{2}{c||}{Selected Decays}  \\
\hline
$k_0$&$K^* (890)$&$ [1^-]$    & 892 ($\pm$), 896(0) &  50      &$^3S_1$&
$K\pi$  &  100\%  \\
\hline
$k_1$&$K_1(1270)$&$  [1^+]$      & 1270        &  90          &$^1P_1$&
$K\rho$  &  42\% \\
&&&&&&
$K^*\pi$  &  16\% \\
&&&&&&
$K\omega$  &  11\% \\
\hline
$k_2$&$K_1(1400)$&$  [1^+]$      & 1402    &  174   &$^3P_1$&
$K\rho$     &   3\%  \\
&&&&&&
$K^*\pi$   &  94\% \\
&&&&&&
$K\omega$ &   1\%   \\
\hline
$k_3$&$K^*(1410)$&$  [1^-]$ & 1412        &  227    &$^3S_1$&
$K\rho$  &$<$ 7\%  \\
&&&&&&
$K^*\pi$   &  $>$ 40\%  \\
&&&&&&
$K\pi$  &  7\%  \\
\hline

$k_4$&$K_2(1430)$&$  [2^+]$  & 1425($\pm$), 1432(0)&  98($\pm$),
109(0)&$^3P_2$& $K\rho$   & 9\%     \\
&&&&&&
$K^*\pi$ & 25\%    \\
&&&&&&
$K\omega$ & 3\%     \\
&&&&&&
$K\pi$  &   50\%  \\
\hline
\end{tabular}
$$

\bigskip\bigskip
Table 1: Some of the properties of the $k_i$ states are shown
\cite{part_data}. Branching fractions of the $k_i$ states to various
final states are given in the last column.
In our
computation, for definiteness,
we used the branching ratios
$Br(k_3\rightarrow K\rho)=7\%$ and
$Br(k_3\rightarrow K^*\pi)=86\%$.

\eject

\bigskip\bigskip

\centerline{\bf Table 2}

$$
\begin{tabular}{||c|c|c||}
\hline
$k_i$   & Model A            &  Model B           \\
\hline
$k_0$   &2.5\%               & 1.6\%              \\
$\hat k_1$&$2.2\times 10^{-5}$ & $1.4\times10^{-5}$ \\
$\hat k_2$&6.5\%               & 1.3\%              \\
$k_3$&3.2\%               & 1.3\%              \\
$k_4$   &5.4\%               & 0.9\%              \\
\hline
\end{tabular}
$$

\bigskip \bigskip

Table 2: The calculated branching fraction for
$B\to k_i\gamma$ in the two models considered.

\bigskip\bigskip \bigskip\bigskip
\bigskip\bigskip \bigskip\bigskip

\centerline{\bf Table 3}

$$
\begin{tabular}{||c|c|c||}
\hline
$k_i$   & Model A            &  Model B           \\
\hline
$k_0$&$ 2.5\times 10^{-3}   $&$ 3.2\times 10^{-3}        $\\
$\hat k_1$&$ -.85   $&$ -.96  $\\
$\hat k_2$&$ 22\times 10^{-3}   $&$ 49\times 10^{-3}         $\\
$k_3$&$ 2.2\times 10^{-3}   $&$ 3.5\times 10^{-3}         $\\
$k_4$&$ 32\times 10^{-3}   $&$  80\times 10^{-3}        $\\
\hline
\end{tabular}
$$

\bigskip \bigskip

Table 3: The resulting ratio of CP
phases, $\sin\phi\over\sin\delta$,
in the two models.

\eject

\centerline{\bf Table 4}
\bigskip\bigskip
\bigskip\bigskip
\centerline{ Model A}

$$
\begin{tabular}{||c|c|c|c||}
\hline
&$\rho K$&$K^* \pi$&$ K\pi$                                \\
\hline
$\fanE_{opt}       $    & 1.0\%   & 1.2\%     &  0.7\%      \\
$\fanE_{0}^\prime  $    & 0.3\%   & 0.3\%     &$1.7\times10^{-4}$   \\
$\fanE_{1}^\prime  $    & 0.7\%   & 0.6\%     &  0.4\%      \\
$\fanE_{2}^\prime  $    & 0.4\%  & 0.2 \%    &                 \\
$    N_{opt}       $    & 33      & 2.2          &  12           \\
$    N_{0}^\prime  $    & 460      & 40        &$2\times 10^4$ \\
$    N_{1}^\prime  $    & 70     & 9       &  32              \\
$    N_{2}^\prime  $    & 280      & 60  &                          \\
\hline
\end{tabular}
$$

\bigskip
\bigskip
\bigskip
\bigskip
\centerline{ Model B}

$$
\begin{tabular}{||c|c|c|c||}
\hline
&$\rho K$&$K^* \pi$&$ K\pi$                                    \\
\hline
$\fanE_{opt}       $    & 2.4\%   & 2.0\%     &  0.9\%          \\
$\fanE_{0}^\prime  $    & 0.6\%   & 0.5\%    &$1\times10^{-4}$  \\
$\fanE_{1}^\prime  $    & 1.7\%   & 1.1\%     &  0.4\%            \\
$\fanE_{2}^\prime  $    & 0.7\%   & 0.3\%     &                   \\
$    N_{opt}       $    & 27      & 3         &  20                 \\
$    N_{0}^\prime  $    & 460      & 47        &$1\times 10^5$        \\
$    N_{1}^\prime  $    & 53      & 10       & $95$                    \\
$    N_{2}^\prime  $    & $310$  &110       &                           \\
\hline
\end{tabular}
$$

\bigskip
\bigskip
\bigskip
\bigskip
Table 4: Using model A and
B we calculate the asymmetries
$\fanE_{opt}$ and $\fanE_i^\prime$ as well as the
corresponding numbers of $B^\pm$ in units of
$10^{8}$. For the annihilation graph we have used the
ISGW model\cite{isgur} to calculate resonance
formation.

\eject

\end{document}